\documentclass[sigconf]{acmart}
\AtBeginDocument{%
  }

\copyrightyear{2025}
\acmYear{2025}
\setcopyright{acmlicensed}
\acmConference[CIKM '25]{Proceedings of the 34th ACM International Conference on Information and Knowledge Management}{November 10--14, 2025}{Seoul, Republic of Korea}
\acmBooktitle{Proceedings of the 34th ACM International Conference on Information and Knowledge Management (CIKM '25), November 10--14, 2025, Seoul, Republic of Korea}
\acmDOI{10.1145/xxxxxxx.xxxxxxx}
\acmISBN{979-8-4007-2040-6/2025/11}

\usepackage{amsmath}
\usepackage{amsfonts}
\usepackage{marvosym}

\usepackage{graphicx}       
\usepackage{booktabs}       
\usepackage{multirow}       
\usepackage{array}          
\usepackage{siunitx}        
\usepackage[table]{xcolor}  

\usepackage{enumitem}       
\usepackage{url}            

\usepackage{color, xspace}  

\begin{document}

\title{SPARK: Adaptive Low-Rank Knowledge Graph Modeling in Hybrid Geometric Spaces for Recommendation
}


\author{Binhao Wang} 
\orcid{0009-0009-6671-3613}
\affiliation{%
  \institution{City University of Hong Kong}
  \city{Hong Kong SAR}
  \country{China}
}
\email{binhawang2-c@my.cityu.edu.hk}

\author{Yutian Xiao} 
\orcid{0000-0002-8276-7920}
\affiliation{%
  \institution{Beihang University}
  \city{Beijing}
  \country{China}
}
\email{by2442221@buaa.edu.cn}

\author{Maolin Wang} 
\orcid{0000-0002-0073-0172}
\affiliation{%
  \institution{City University of Hong Kong}
  \city{Hong Kong SAR}
  \country{China}
}
\email{MorinWang@foxmail.com}

\author{Zhiqi Li}
\orcid{0009-0003-5004-8213}
\affiliation{%
  \institution{City University of Hong Kong}
  \city{Hong Kong SAR}
  \country{China}
}
\email{zhiqli6-c@my.cityu.edu.hk}

\author{Tianshuo Wei}
\orcid{0009-0008-2470-6688}
\affiliation{%
  \institution{City University of Hong Kong}
  \city{Hong Kong SAR}
  \country{China}
}
\email{tianshwei2-c@my.cityu.edu.hk}

\author{Ruocheng Guo}
\orcid{0000-0002-8522-6142}
\affiliation{%
  \institution{Independent Researcher}
  \city{Hong Kong SAR}
  \country{China}
}
\email{rguo.asu@gmail.com}

\author{Xiangyu Zhao} 
\orcid{0000-0003-2926-4416}
\authornote{Corresponding author.}
\affiliation{%
  \institution{City University of Hong Kong}
  \city{Hong Kong SAR}
  \country{China}
}
\email{xianzhao@cityu.edu.hk}

%
\renewcommand{\shortauthors}{Binhao Wang et al.}

\begin{abstract}
Knowledge Graphs (KGs) enhance recommender systems but face challenges from inherent noise, sparsity, and Euclidean geometry's inadequacy for complex relational structures, critically impairing representation learning, especially for long-tail entities. Existing methods also often lack adaptive multi-source signal fusion tailored to item popularity.
This paper introduces SPARK, a novel multi-stage framework systematically tackling these issues. SPARK first employs Tucker low-rank decomposition to denoise KGs and generate robust entity representations. Subsequently, an SVD-initialized hybrid geometric GNN concurrently learns representations in Euclidean and Hyperbolic spaces; the latter is strategically leveraged for its aptitude in modeling hierarchical structures, effectively capturing semantic features of sparse, long-tail items. A core contribution is an item popularity-aware adaptive fusion strategy that dynamically weights signals from collaborative filtering, refined KG embeddings, and diverse geometric spaces for precise modeling of both mainstream and long-tail items. Finally, contrastive learning aligns these multi-source representations.
Extensive experiments demonstrate SPARK's significant superiority over state-of-the-art methods, particularly in improving long-tail item recommendation, offering a robust, principled approach to knowledge-enhanced recommendation. Implementation code is anonymously online \footnote{ \url{https://github.com/Applied-Machine-Learning-Lab/SPARK}}.
\end{abstract}




\maketitle

\section{Introduction}
\label{sec:introduction}

\begin{figure}[htbp]
	\centering
	\includegraphics[width=1.0\columnwidth]{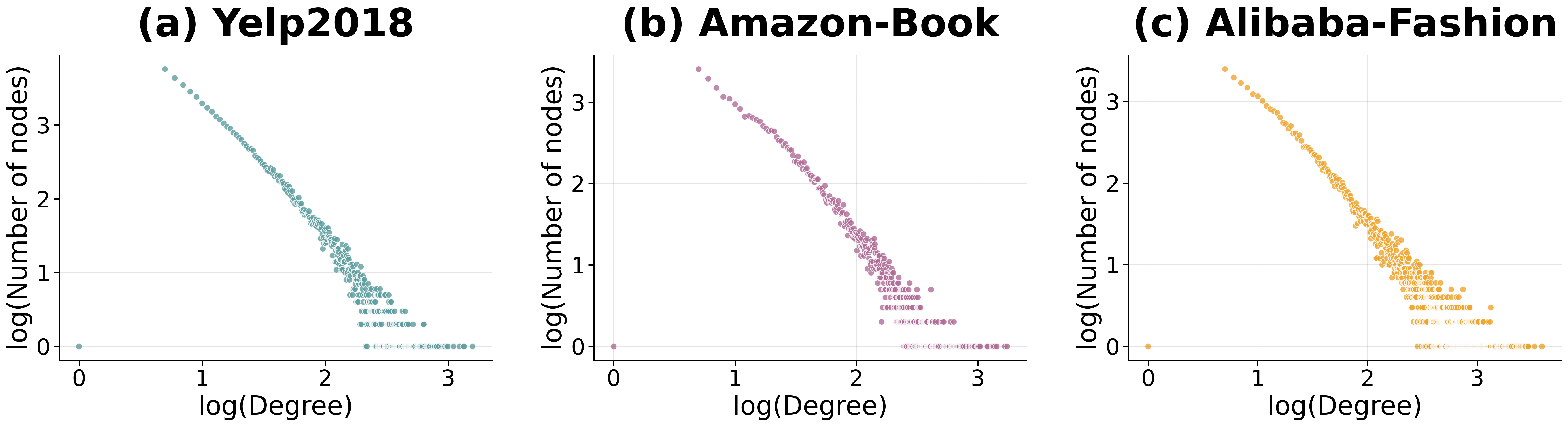}
	\caption{Degree distribution across different datasets: (a) Yelp2018, (b) Amazon-Book, and (c) Alibaba-Fashion. All datasets exhibit the characteristic long-tail distribution.}
	\label{fig_degree_distribution}

    \vspace{-3mm}
\end{figure}

Knowledge Graphs (KGs)~\cite{hogan2021knowledge} are pivotal for enhancing recommender systems by providing rich semantic context, alleviating data sparsity and cold-start issues~\cite{guo2020survey, lin2023autodenoise}. However, effectively leveraging KGs remains challenging. 
Firstly, as shown in Figure~\ref{fig_degree_distribution}, real-world KGs often suffer from noise and exhibit pronounced long-tail distributions. This characteristic, where most entities have few connections~\cite{wang2019tackling, sun2020multi}, hinders robust representation learning, especially for low-frequency entities~\cite{celma2008hits, liu2023exploration}. 
Secondly, conventional Euclidean geometry struggles to capture complex hierarchical or power-law structures~\cite{bronstein2021geometric, chami2019hyperbolic} in KGs and user-item interactions, limiting representational capacity~\cite{nickel2018learning}. 
Finally, adaptively fusing diverse signals—collaborative, KG-semantic, and geometric—for mainstream and long-tail items remains a significant hurdle~\cite{zhang2022multitask, li2023hamur, fu2025unified}.

While Graph Neural Networks (GNNs)~\cite{wu2022survey} and hyperbolic geometry explorations~\cite{chamberlain2019scalable, liu2019hyperbolic} show promise in integrating KG information or capturing hierarchies, existing methods often inadequately address the interplay of these challenges, particularly for long-tail item recommendation. Many approaches are susceptible to KG noise affecting sparse entities~\cite{lin2015learning, sun2019rotate}, or their Euclidean GNNs fail to model the nuanced semantics crucial for long-tail items. Furthermore, simplistic or static fusion strategies may not effectively leverage diverse signals for items across varying popularity levels~\cite{abdollahpouri2020popularity}, neglecting the unique requirements of long-tail items that could significantly benefit from robust knowledge graph semantics~\cite{zhang2019long} or hyperbolic geometric priors~\cite{Apanasov1997,yang2022hrcf}.

To holistically address these limitations, we propose \textbf{SPARK} (\textbf{S}elf-adaptive \textbf{P}opularity-\textbf{A}ware \textbf{R}ank-decomposition for \textbf{K}nowledge-enhanced Recommendation). SPARK features a multi-stage pipeline designed to enhance noise robustness and fortify long-tail item representation.
(1) For robust KG preprocessing, it employs low-rank Tucker decomposition~\cite{balazevic2019tucker, zhang2020deep, xu2024multi} to denoise KGs and generate robust initial entity representations, which is vital for the long tail.
(2) For hybrid geometric learning, an SVD-initialized hybrid GNN~\cite{liu2019hyperbolic, peng2022svd, zhang2021lorentzian} concurrently learns Euclidean and Hyperbolic representations~\cite{ju2024hypergraph, ju2024comprehensive, ganea2018hyperbolic}. The hyperbolic space is strategically used to effectively model complex structures~\cite{bronstein2017geometric, krioukov2010hyperbolic} and enhance the semantic understanding of sparse, long-tail items~\cite{chamberlain2017neural}.
(3) For adaptive multi-signal fusion, an item popularity-aware mechanism~\cite{wei2021model, yin2012challenging} dynamically weights collaborative~\cite{wang2019kgat, kim2020multi}, KG-derived, and geometric signals~\cite{chami2019hyperbolic, liu2019hyperbolic} for precise, differentiated modeling of mainstream and long-tail items~\cite{li2023hamur, fu2025unified}.
(4) Finally, for contrastive alignment, contrastive learning aligns these multi-source signals~\cite{xie2022self, yang2022knowledge}, improving representation consistency and generalization~\cite{liu2021contrastive, qiu2020gcc}.

The primary contributions of this paper are summarized as follows:
\begin{itemize}[leftmargin=*]
    \item \textbf{Novel Framework for Robust Knowledge-Enhanced Recommendation.} We propose \textbf{SPARK}, a multi-stage framework that systematically addresses critical challenges in knowledge-aware recommendation. SPARK integrates (1) low-rank Tucker decomposition for KG preprocessing to mitigate noise and enhance entity representations, especially for long-tail items, and (2) a popularity-aware adaptive fusion mechanism to dynamically balance diverse information sources.

    \item \textbf{Hybrid Geometric Modeling for Enhanced Representation Learning.} We design a hybrid geometric Graph Neural Network (GNN) architecture that operates on SVD-initialized features. This architecture concurrently learns representations in both Euclidean and Hyperbolic spaces. The hyperbolic component is particularly leveraged for its inherent strength in modeling hierarchical structures and complex relationships, thereby improving the semantic understanding and representation quality of items, especially those in the long tail. This dual-space approach allows SPARK to overcome limitations of purely Euclidean methods.

    \item \textbf{Empirical Validation and Superior Performance.} Extensive experiments conducted on three real-world benchmark datasets demonstrate the effectiveness of SPARK. Our results show that SPARK significantly outperforms various state-of-the-art baselines, particularly highlighting its capability to improve recommendation quality for long-tail items. Ablation studies further verify the importance and rationality of each key component within the SPARK framework.
\end{itemize}

\begin{figure*}[htbp]
    \centering
    \includegraphics[width=0.95\linewidth]{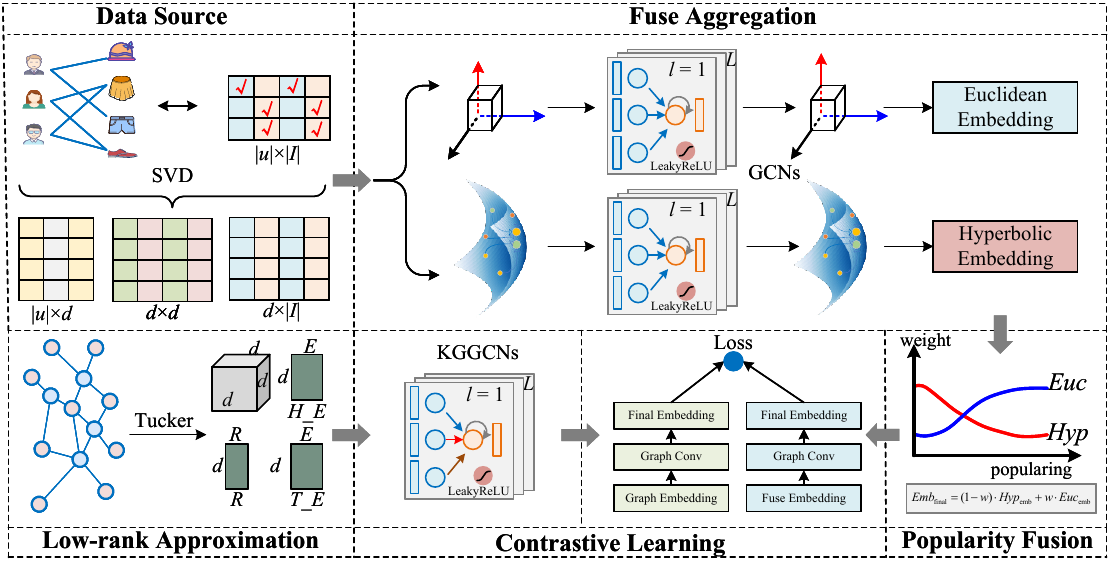} 
    \caption{The overall architecture of  {SPARK}. The framework integrates (1) Knowledge Graph (KG) representation learning with Tucker-based scoring for KG triples; (2) SVD-initialized Hybrid Geometric GNNs operating in parallel Euclidean and Hyperbolic (Lorentz model) spaces on the user-item interaction graph; (3) Contrastive alignment of SVD-derived collaborative views and KG-enhanced semantic views; and (4) Popularity-aware adaptive fusion to dynamically combine signals for final recommendation.}
    \label{fig:main_architecture_spark_hygrec_style_full}
\end{figure*}

\section{Preliminaries}\label{sec:preliminaries}

This section introduces essential background on hyperbolic geometry, knowledge graphs, and the formal problem definition for knowledge-aware recommendation, setting the stage for our proposed framework.

\subsection{Hyperbolic Geometry}\label{subsec:hyperbolic_geometry}

We employ hyperbolic geometry, specifically the Lorentz model~\cite{nickel2018learning,ganea2018hyperbolic}, for its efficacy in modeling hierarchical data. Among various hyperbolic models, we choose the Lorentz model for its numerical stability and the straightforward formulation of essential geometric operations, which simplifies the implementation of complex GNN layers compared to other models like the Poincaré ball~\cite{liu2019hyperbolic}.
A $d$-dimensional hyperbolic space is represented as the manifold $\mathbb{L}^{d,c}$, defined as:
\begin{equation}\label{eq:lorentz_manifold}
\mathbb{L}^{d,c} = \{ \mathbf{x} \in \mathbb{R}^{d+1} \,|\, \langle \mathbf{x}, \mathbf{x} \rangle_{\mathcal{L}} = -c, \text{ and } x_0 > 0 \},
\end{equation}
which has a constant negative curvature of $-\frac{1}{c}$ (where $c > 0$). The Lorentzian inner product for two vectors $\mathbf{a}, \mathbf{b} \in \mathbb{R}^{d+1}$ is given by $\langle \mathbf{a}, \mathbf{b} \rangle_{\mathcal{L}} = -a_0 b_0 + \sum_{i=1}^{d} a_i b_i$.
The tangent space at a point $\mathbf{x} \in \mathbb{L}^{d,c}$ is denoted $\mathbb{T}_{\mathbf{x}}^{d,c}$ and defined as:
\begin{equation}\label{eq:tangent_space}
\mathbb{T}_{\mathbf{x}}^{d,c} = \{ \mathbf{v} \in \mathbb{R}^{d+1} \,|\, \langle \mathbf{v}, \mathbf{x} \rangle_{\mathcal{L}} = 0 \}.
\end{equation}
Within this tangent space, the Riemannian norm of a vector $\mathbf{v}$ is $\|\mathbf{v}\|_{\mathbf{x}} = \sqrt{\langle \mathbf{v}, \mathbf{v} \rangle_{\mathbf{x}}}$, where $\langle \cdot, \cdot \rangle_{\mathbf{x}}$ is the induced metric on the tangent space (often the same as $\langle \cdot, \cdot \rangle_{\mathcal{L}}$ when restricted).

The exponential map $\exp_{\mathbf{x}} : \mathbb{T}_{\mathbf{x}}^{d,c} \to \mathbb{L}^{d,c}$ projects a tangent vector $\mathbf{v}$ from the tangent space at $\mathbf{x}$ onto the manifold:
\begin{equation}\label{eq:exp_map_corrected}
\exp_{\mathbf{x}}(\mathbf{v}) = \begin{cases}
\cosh\left(\frac{\|\mathbf{v}\|_{\mathbf{x}}}{\sqrt{c}}\right)\mathbf{x} + \sqrt{c} \sinh\left(\frac{\|\mathbf{v}\|_{\mathbf{x}}}{\sqrt{c}}\right) \frac{\mathbf{v}}{\|\mathbf{v}\|_{\mathbf{x}}}, & \text{if } \|\mathbf{v}\|_{\mathbf{x}} > 0 \\
\mathbf{x}, & \text{if } \|\mathbf{v}\|_{\mathbf{x}} = 0.
\end{cases}
\end{equation}
Conversely, the logarithmic map $\log_{\mathbf{x}} : \mathbb{L}^{d,c} \to \mathbb{T}_{\mathbf{x}}^{d,c}$ maps a point $\mathbf{y}$ on the manifold back to the tangent space at $\mathbf{x}$:
\begin{equation}\label{eq:log_map_corrected}
\log_{\mathbf{x}}(\mathbf{y}) = \begin{cases}
\frac{d_{\mathbb{L}}(\mathbf{x}, \mathbf{y})}{\|\mathbf{p}_{\mathbf{x}\to\mathbf{y}}\|_{\mathbf{x}}} \mathbf{p}_{\mathbf{x}\to\mathbf{y}}, & \text{if } \|\mathbf{p}_{\mathbf{x}\to\mathbf{y}}\|_{\mathbf{x}} > 0 \\
\mathbf{0}, & \text{if } \|\mathbf{p}_{\mathbf{x}\to\mathbf{y}}\|_{\mathbf{x}} = 0,
\end{cases}
\end{equation}
where $\mathbf{p}_{\mathbf{x}\to\mathbf{y}} = \mathbf{y} + \frac{\langle \mathbf{x}, \mathbf{y} \rangle_{\mathcal{L}}}{c} \mathbf{x}$ is the projection of $\mathbf{y}$ onto the tangent space after parallel transport considerations (this definition of $\mathbf{p}_{\mathbf{x}\to\mathbf{y}}$ is specific to the Lorentz model and ensures $\log_{\mathbf{x}}(\mathbf{y})$ is in $\mathbb{T}_{\mathbf{x}}^{d,c}$).
Finally, the geodesic distance between two points $\mathbf{x}, \mathbf{y} \in \mathbb{L}^{d,c}$ is calculated as:
\begin{equation}\label{eq:geodesic_distance}
d_{\mathbb{L}}(\mathbf{x}, \mathbf{y}) = \sqrt{c} \operatorname{arccosh}\left(\max\left(1, -\frac{1}{c}\langle \mathbf{x}, \mathbf{y} \rangle_{\mathcal{L}}\right)\right).
\end{equation}

\subsection{Knowledge Graph}\label{subsec:knowledge_graph}
A Knowledge Graph (KG) is formally defined as $\mathcal{G}_{KG} = \{ (h, r, t) | h, t \in \mathcal{E}, r \in \mathcal{R} \}$, where $\mathcal{E}$ is the set of entities and $\mathcal{R}$ the set of relations. These (head, relation, tail) triples form a multi-relational graph capturing rich semantic connections. While KGs provide substantial semantic context, their structural complexity, large scale, and often noisy nature present challenges for effective modeling and utilization. KGs are particularly crucial for enriching item representations and alleviating data sparsity for long-tail items.

\subsection{Problem Definition}\label{subsec:problem_definition}
Given a set of users $\mathcal{U}$, a set of items $\mathcal{I}$ (where items are a subset of KG entities, $\mathcal{I} \subseteq \mathcal{E}$), their implicit interactions $\mathbf{R} \in \{0, 1\}^{|\mathcal{U}| \times |\mathcal{I}|}$, and a knowledge graph $\mathcal{G}_{KG}$, the goal of knowledge-aware recommendation is to predict a personalized ranked list of unobserved items for each user $u \in \mathcal{U}$. This involves learning a scoring function $\hat{y}_{ui} = f(u, i | \mathbf{R}, \mathcal{G}_{KG})$ that estimates user $u$'s preference for item $i$. Our framework aims to learn robust user and item representations by effectively integrating $\mathbf{R}$ and $\mathcal{G}_{KG}$, while addressing challenges such as KG noise and long-tail distributions.

\section{Method}
\label{sec:method_hygrec_style_full_balanced}

We introduce SPARK, a multi-stage framework designed to enhance knowledge-aware recommendation. It systematically addresses challenges such as KG noise, inadequate long-tail item representation, and the limitations of conventional geometric spaces. The architecture of SPARK, illustrated in Figure~\ref{fig:main_architecture_spark_hygrec_style_full}, integrates several key components detailed below.

\subsection{KG Processing with Low-Rank Tucker Decomposition}
\label{subsec:kg_processing_tucker_balanced}
Knowledge Graphs (KGs) provide rich semantic context that extends beyond user-item interaction data. However, raw KGs can be noisy and voluminous. To effectively harness this information, SPARK first processes the input KG to distill robust semantic representations. We maintain learnable embeddings for all entities $\mathcal{E}$ and relations $\mathcal{R}$ within the KG, denoted as $\mathbf{E}_{\text{ent}} \in \mathbb{R}^{|\mathcal{E}| \times d}$ and $\mathbf{E}_{\text{rel}} \in \mathbb{R}^{|\mathcal{R}| \times d}$ respectively, where $d$ is their dimensionality. These embeddings are initialized and refined during training.

To explicitly model the multi-way interactions inherent within KG triples $(h,r,t)$ and crucially, to generate more compact and noise-resilient semantic representations, particularly for less frequent long-tail entities, we strategically employ a scoring function inspired by low-rank Tucker decomposition, as proposed in TuckER~\cite{balazevic2019tucker}. Compared to other tensor factorization methods like CP decomposition, Tucker's use of a core tensor provides greater modeling capacity to capture the complex, multi-way interactions within KG triples, making it particularly effective for denoising and generating robust representations~\cite{nickel2017holographic}. This captures complex multi-linear relationships effectively. The initial $d$-dimensional entity and relation embeddings ($\mathbf{e}_h, \mathbf{e}_r, \mathbf{e}_t$) are transformed via learnable matrices $\mathbf{W}_e, \mathbf{W}_r \in \mathbb{R}^{d \times d_c}$ into a shared core latent space of dimension $d_c$, yielding $\mathbf{h}', \mathbf{r}', \mathbf{t}'$. The plausibility score of a triple is then computed as:
\vspace{-0.5mm}
\begin{equation}\label{eq:tucker_score_spark_hygrec_full_balanced}
s(h,r,t) = \langle \psi(\text{BN}(\mathbf{h}'), \text{BN}(\mathbf{r}'); \mathcal{W}), \text{BN}(\mathbf{t}') \rangle,
\end{equation}
\vspace{-0.5mm}
where $\mathcal{W} \in \mathbb{R}^{d_c \times d_c \times d_c}$ is a learnable core tensor central to modeling interactions. The function $\psi(\cdot)$ encapsulates the multi-linear product (e.g., $\mathbf{h}' (\mathbf{r}'\mathcal{W})$ after appropriate tensor reshaping). Batch Normalization (BN) and dropout are applied for regularization. This Tucker-based scoring mechanism is integrated into a BPR-like pairwise ranking loss, $\mathcal{L}_{TuckER}$, which encourages valid KG triples to receive higher scores than corrupted ones, guiding the learning of high-quality, noise-robust KG embeddings.

\subsection{SVD-Initialized Hybrid Geometric Graph Neural Networks}
\label{subsec:hybrid_gnn_hygrec_style_full_balanced}
SPARK models multifaceted user preferences by synergizing global collaborative trends, distilled from the entire user-item interaction matrix $\mathbf{R}$ via Singular Value Decomposition (SVD), with local interaction patterns captured from neighborhood structures by Graph Neural Networks (GNNs). This synergy is realized through an SVD-initialized hybrid geometric GNN framework.
Recognizing that relational data often exhibits characteristics ill-suited for purely Euclidean spaces (e.g., hierarchies), our framework further incorporates both Euclidean and Hyperbolic geometric pathways to enhance representational power.

User and item features are initialized using pre-computed SVD components from $\mathbf{R}$. The top-$k_s$ singular vectors for users ($\mathbf{U}_s \in \mathbb{R}^{M \times k_s}$) and items ($\mathbf{V}_s \in \mathbb{R}^{N \times k_s}$), along with singular values ($\mathbf{\Sigma}_s \in \mathbb{R}^{k_s \times k_s}$), are utilized. An exponential filter $f(\sigma_j) = e^{\beta \sigma_j}$ is applied to $\mathbf{\Sigma}_s$ to emphasize significant collaborative patterns. The initial Euclidean features are then:
\begin{align}
\mathbf{H}_{U,E}^{(0)} &= \mathbf{U}_s \text{diag}(f(\mathbf{\Sigma}_s)) \mathbf{F}_S, \label{eq:svd_user_init_balanced} \\
\mathbf{H}_{I,E}^{(0)} &= \mathbf{V}_s \text{diag}(f(\mathbf{\Sigma}_s)) \mathbf{F}_S, \label{eq:svd_item_init_balanced}
\end{align}
where $\mathbf{F}_S \in \mathbb{R}^{k_s \times d_e}$ is a learnable projection matrix mapping SVD features to the GNN embedding dimension $d_e$. These SVD-initialized embeddings provide a robust global collaborative prior, bootstrapping GNN learning.

These embeddings $(\mathbf{H}_{U,E}^{(0)}, \mathbf{H}_{I,E}^{(0)})$ are the common input to two parallel GNN pathways operating on the user-item interaction graph $\mathcal{G}_{UI}$ (normalized adjacency matrix $\mathbf{A}_{\text{UI}}$).

\textbf{Euclidean GNN Pathway:} This pathway refines $\mathbf{H}_{E}^{(0)}$ over $L$ layers using standard graph convolutions. At each layer $l$:
\begin{equation}\label{eq:euclidean_gnn_spark_hygrec_full_balanced}
\mathbf{H}_{E}^{(l)} = \text{Activate} \left( \alpha_E \mathbf{A}_{\text{UI}} \mathbf{H}_{E}^{(l-1)} \mathbf{W}_{E}^{(l)} + (1-\alpha_E) \mathbf{H}_{E}^{(l-1)} \mathbf{W}_{E,self}^{(l)} \right),
\end{equation}
where $\mathbf{W}_{E}^{(l)}, \mathbf{W}_{E,self}^{(l)}$ are learnable weights and $\alpha_E$ balances their contributions. This yields refined Euclidean embeddings $(\mathbf{H}_{U,E}^{(L)}, \mathbf{H}_{I,E}^{(L)})$, adept at capturing standard collaborative patterns.

\textbf{Hyperbolic GNN Pathway (Lorentz Model):} To leverage hyperbolic geometry's capacity for embedding hierarchical structures, initial Euclidean features $\mathbf{H}_{E}^{(0)}$ are mapped to initial hyperbolic embeddings $\mathbf{H}_{H}^{(0)}$ on the Lorentz manifold $\mathbb{L}^{d_e+1, c}$ (as detailed in \S\ref{subsec:hyperbolic_geometry}, typically by projecting to $\mathcal{T}_{\mathbf{o}}\mathbb{L}$ then applying $\exp_{\mathbf{o}}(\cdot)$). These $\mathbf{H}_{H}^{(0)}$ are refined over $L$ GNN layers. Operations respect hyperbolic geometry: embeddings are mapped from $\mathbb{L}$ to $\mathcal{T}_{\mathbf{o}}\mathbb{L}$ via $\log_{\mathbf{o}}(\cdot)$; aggregation and learnable transformations occur in this tangent space; results are mapped back to $\mathbb{L}$ via $\exp_{\mathbf{o}}(\cdot)$. Möbius addition $\oplus_M$ and scalar multiplication $\otimes_M$ are used for residual connections and geometric integrity. The layer $l$ update is:
\begin{equation}
\label{eq:hyperbolic_gnn_spark_hygrec_full_balanced_split}
\begin{split}
\mathbf{H}_{H}^{(l)} ={}& (w_s \otimes_M \exp_{\mathbf{o}}(\text{Agg}_{\text{hyp}}^{(l)}(\log_{\mathbf{o}}(\mathbf{H}_{H}^{(l-1)})))) \\
&\oplus_M ((1-w_s) \otimes_M \mathbf{H}_{H}^{(l-1)}),
\end{split}
\end{equation}
where $\text{Agg}_{\text{hyp}}^{(l)}$ is the hyperbolic aggregation layer and $w_s$ a skip-connection weight, producing $(\mathbf{H}_{U,H}^{(L)}, \mathbf{H}_{I,H}^{(L)})$. These are effective for items with implicit hierarchies or complex relational paths. An optional final adjustment can ensure embeddings strictly reside on the manifold.

\subsection{Contrastive View Alignment}\label{subsec:contrastive_alignment_balanced}
To foster robust item representations, SPARK aligns views from different perspectives: a global SVD-Collaborative View ($\mathbf{z}_{i,SVD}$) and a KG-Enhanced Semantic View ($\mathbf{z}_{i,KG}$). This encourages consistency and complementarity. The alignment operates on pairs of augmented views for each item $i$ in a training batch $\mathcal{B}$.

The SVD-Collaborative View $\mathbf{z}_{i,SVD}$ derives from refined Euclidean GNN item embeddings $\mathbf{H}_{I,E}^{(L)}$, passed through projection head $f_{SVD}(\cdot)$.
The KG-Enhanced Semantic View $\mathbf{z}_{i,KG}$ originates from the item's KG entity embedding $\mathbf{e}_i$ (\S\ref{subsec:kg_processing_tucker_balanced}). This $\mathbf{e}_i$ is processed by a KG-specific graph aggregator (`KGGCNNs` in Figure~\ref{fig:main_architecture_spark_hygrec_style_full}) on the KG's structure, then passed through projection $f_{KG}(\cdot)$.

Both L2-normalized projected views are aligned via bidirectional InfoNCE loss, maximizing agreement between positive pairs (same item, different views) while distinguishing them from negative pairs:
\begin{equation}\label{eq:contrastive_loss_spark_hygrec_full_balanced}
\begin{split}
\mathcal{L}_{CL} = -\frac{1}{|\mathcal{B}|}\sum_{i \in \mathcal{B}} \Biggl( &\log \frac{\exp(\text{sim}(\mathbf{z}_{i,SVD}, \mathbf{z}_{i,KG})/\tau)}{\sum_{j \in \mathcal{B}} \exp(\text{sim}(\mathbf{z}_{i,SVD}, \mathbf{z}_{j,KG})/\tau)} \\
&+ \log \frac{\exp(\text{sim}(\mathbf{z}_{i,KG}, \mathbf{z}_{i,SVD})/\tau)}{\sum_{j \in \mathcal{B}} \exp(\text{sim}(\mathbf{z}_{i,KG}, \mathbf{z}_{j,SVD})/\tau)} \Biggr),
\end{split}
\end{equation}
where $\text{sim}(\cdot,\cdot)$ is cosine similarity and $\tau$ a temperature hyperparameter. This encourages learning view-invariant representations.

\subsection{Popularity-Aware Adaptive Fusion}\label{subsec:adaptive_fusion_balanced_concise}
To handle varied item popularity and signal reliability, SPARK employs an adaptive fusion strategy. Based on item popularity $p_i$ (e.g., from interaction frequency), a dynamic gating weight $w'_i \in [0,1]$ is computed:
\begin{equation}\label{eq:fusion_weight_actual_concise}
w'_i = \sigma_{\text{gate}}(p_i; \theta_{\text{gate}}),
\end{equation}
where $\sigma_{\text{gate}}(\cdot)$ is a parameterized sigmoid, yielding larger $w'_i$ for more popular items.

This weight $w'_i$ tailors the final item representations. The final Euclidean item representation, $\mathbf{H}_{I,E}^{\text{final}}$, blends collaborative signals and KG semantics:
\begin{equation}\label{eq:final_euclidean_item_emb_actual_concise}
\mathbf{H}_{I,E}^{\text{final}} = w'_i \cdot \mathbf{H}_{I,E}^{(L)} + (1-w'_i) \cdot \mathbf{H}_{I,E}^{\text{fused}}.
\end{equation}
Here, $\mathbf{H}_{I,E}^{(L)}$ is the Euclidean GNN output. The KG-synergized embedding, $\mathbf{H}_{I,E}^{\text{fused}}$, aims to enrich sparser items. It is formed by: (1) an SVD-guided attention $\text{Attn}(\mathbf{H}_{I,E}^{(0)}, \mathbf{e}_{i,KG}^{\text{agg}})$ on the KG-aggregated embedding $\mathbf{e}_{i,KG}^{\text{agg}}$ (from `KGGCNNs` in Figure~\ref{fig:main_architecture_spark_hygrec_style_full}) to get $\mathbf{e}_{i,KG}^{\text{att}}$; (2) combining $\mathbf{e}_{i,KG}^{\text{att}}$ with $\mathbf{H}_{I,E}^{(L)}$ (e.g., via concatenation) and processing through a fusion MLP, $f_{\text{fuse}}(\cdot)$, to yield $\mathbf{H}_{I,E}^{\text{fused}}$.

Subsequently, the final Hyperbolic item representation $\mathbf{H}_{I,H}^{\text{final}}$ results from mapping $\mathbf{H}_{I,E}^{\text{final}}$ to the Lorentz manifold, allowing the hyperbolic component to benefit from this adaptive KG infusion. User representations $(\mathbf{H}_{U,E}^{\text{final}}, \mathbf{H}_{U,H}^{\text{final}})$ are typically their GNN outputs $(\mathbf{H}_{U,E}^{(L)}, \mathbf{H}_{U,H}^{(L)})$.

The predicted preference score $\hat{y}_{ui}$ adaptively combines signals from both geometric spaces, modulated by $w'_i$:
\begin{equation}\label{eq:rec_score_spark_hygrec_full_balanced_no_bold_concise}
\hat{y}_{ui} = w'_i \cdot \langle \mathbf{H}_{U,E}^{\text{final}}, \mathbf{H}_{I,E}^{\text{final}} \rangle + (1-w'_i) \cdot (-d_{\mathbb{L}}(\mathbf{H}_{U,H}^{\text{final}}, \mathbf{H}_{I,H}^{\text{final}})).
\end{equation}
This allows dynamic emphasis on Euclidean (for popular items) or KG-enriched Hyperbolic perspectives (for long-tail items). The Tucker-based KG score $s(u, \text{interacts}, i)$ can be additively integrated for explicit KG knowledge.

\subsection{Overall Training Objective}\label{subsec:training_objective_balanced}
SPARK is trained end-to-end by optimizing a composite loss:
\begin{equation}\label{eq:total_loss_spark_hygrec_full_balanced_no_bold}
\mathcal{L} = \mathcal{L}_{\text{REC}} + \lambda_{CL} \mathcal{L}_{CL} + \lambda_{TuckER} \mathcal{L}_{TuckER} + \lambda_{REG} \mathcal{L}_{REG}.
\end{equation}
Here, $\mathcal{L}_{\text{REC}}$ is the primary recommendation loss (e.g., BPR loss~\cite{rendle2009bpr} on $\hat{y}_{ui}$). $\mathcal{L}_{CL}$ is the contrastive alignment loss (Eq.~\ref{eq:contrastive_loss_spark_hygrec_full_balanced}). $\mathcal{L}_{TuckER}$ is the BPR-like loss for KG triple scoring from \S\ref{subsec:kg_processing_tucker_balanced}. $\mathcal{L}_{REG}$ is an L2 regularization term. Hyperparameters $\lambda_{CL}, \lambda_{TuckER}, \lambda_{REG}$ balance these components.

\section{Experiments}
\label{sec:experiments}

In this section, we conduct a series of experiments to validate SPARK and answer the following key research questions:
\begin{itemize}[leftmargin=*]
    \item\textbf{RQ1}: How does SPARK perform compared to baseline models on real-world datasets?
    \item\textbf{RQ2}: How does each proposed module of SPARK contribute to the overall performance?
    \item\textbf{RQ3}: How do hyperparameters influence the performance of SPARK?
    \item\textbf{RQ4}: How effectively does SPARK address the long-tail item problem through improved embedding distributions?
\end{itemize}

\subsection{Experimental Settings}

\subsubsection{\textbf{Datasets and Metrics}}
To rigorously evaluate the effectiveness of SPARK, we employ three widely used real-world datasets, chosen for their diverse characteristics and common use in recommendation research: Amazon-Book\footnote{\url{https://jmcauley.ucsd.edu/data/amazon/}} (rich in textual reviews and item metadata), Alibaba-iFashion\footnote{\url{https://github.com/wenyuer/POG}} (representing a large-scale e-commerce scenario), and Yelp2018\footnote{\url{https://www.yelp.com/dataset/challenge}} (featuring interactions with local businesses and social connections). The statistics of these datasets, including user-item interactions and knowledge graph properties, are summarized in Table~\ref{tab:dataset_stats_spark_right_align}. Consistent with previous studies \cite{wang2019kgat}, knowledge graphs for Yelp2018 and Amazon-Book are constructed using their established methodologies. For the Alibaba-iFashion dataset, we follow the knowledge graph construction protocol introduced in \cite{wang2019knowledge}. Each dataset is partitioned into training, validation, and test subsets with a standard 8:1:1 ratio to ensure robust evaluation.

\begin{table}[!t] 
\centering
\small
\caption{Statistics of the datasets used for evaluation, detailing user-item interactions and knowledge graph properties.}
\label{tab:dataset_stats_spark_right_align}
\setlength{\tabcolsep}{4pt}
\begin{tabular}{@{}ll rrr@{}}
\toprule
 & Metric & Amazon-Book & Alibaba-iFashion & Yelp2018 \\
\midrule
\multirow{4}{*}{\begin{tabular}[c]{@{}l@{}}User-Item\\Interaction\end{tabular}}
& \#Users      & 70,679   & 114,737  & 45,919   \\
& \#Items      & 24,915   & 30,040   & 45,538   \\
& \#Interactions & 847,733  & 1,781,093 & 1,185,068 \\
& Sparsity (\%)& 99.952  & 99.948  & 99.943  \\
\midrule
\multirow{3}{*}{\begin{tabular}[c]{@{}l@{}}Knowledge\\Graph\end{tabular}}
& \#Entities   & 88,572   & 59,156   & 90,961   \\
& \#Relations  & 39      & 51      & 42      \\
& \#Triplets   & 2,557,746 & 279,155  & 1,853,704 \\
\bottomrule
\end{tabular}
\end{table}

We adopt two widely recognized evaluation metrics, Recall@N~\cite{elliott2024impacts} and NDCG@N~\cite{wang2023theoretical}, to assess the top-$N$ recommendation performance, as they effectively measure both the retrieval of relevant items and the quality of their ranking. Specifically, we set $N$ values to 10 and 20 by default (denoted as R@10, N@10, R@20, N@20) and report the average performance over all users in the test set.

\subsubsection{\textbf{Baselines and Hyperparameter Settings}}

The effectiveness of SPARK is assessed through comparison with a comprehensive suite of baselines. These include: classic collaborative filtering methods like LightGCN \cite{he2020lightgcn} and SGL \cite{wu2021self}; SVD-enhanced approaches such as SVD-GCN \cite{peng2022svd}; methods leveraging hyperbolic geometry, namely HGCF \cite{sun2021hgcf} and HICF \cite{yang2022hicf}; and prominent knowledge-aware techniques including KGCN \cite{wang2019knowledge}, KGIN \cite{wang2021learning}, KGAT \cite{wang2019kgat}, KGCL \cite{yang2022knowledge}, LKGR \cite{chen2022modeling}, and HCMKR \cite{sun2024hyperbolic}. This diverse set of baselines allows for a thorough evaluation of SPARK's capabilities against various established paradigms. Unlike baselines that typically address single aspects of the recommendation problem, SPARK offers a holistic solution by comprehensively combining SVD initialization, hybrid geometric modeling, and principled knowledge graph processing within a unified framework.

The SPARK model is implemented in Python 3.10 using PyTorch 2.1.0. For fair comparison, baseline methods are evaluated using the RecBole \cite{recbole[1.0]} library with uniform settings: embedding dimensionality is set to 64, batch size to 1024, and the Adam optimizer~\cite{kingma2014adam} is used for its efficiency and adaptive learning rate capabilities. For a fair comparison in embedding dimensionality, we clarify that SPARK does not concatenate its Euclidean and Hyperbolic embeddings. Hyperparameters for all models, including SPARK, are optimized via grid search on the validation set, with learning rates explored in $\{10^{-4}, 2\times10^{-4}, 5\times10^{-4}\}$ and GNN layers in $\{1,2,3,4\}$. Key hyperparameter choices for SPARK, such as SVD embedding dimension, Tucker core dimension, and GNN layers, are further analyzed in Section~\ref{subsec:hyperparam_sensitivity_spark}. Experiments are conducted on a Linux server equipped with an NVIDIA RTX 4090 GPU.

\begin{table*}[htbp]
\centering
\small
\caption{Overall performance comparison on three datasets. Best results are in \textbf{bold}, second best are \underline{underlined}. * denotes statistical significance ($p < 0.05$) over the best baseline.}
\label{tab:overall_results_spark} 
\renewcommand{\arraystretch}{1.1}
\setlength{\tabcolsep}{1.5mm} 
\begin{tabular}{@{}l cccc cccc cccc@{}}
 \toprule
 \multirow{2}{*}{Model} & \multicolumn{4}{c}{Amazon-Book} & \multicolumn{4}{c}{Alibaba-iFashion} & \multicolumn{4}{c}{Yelp2018} \\
 \cmidrule(lr){2-5} \cmidrule(lr){6-9} \cmidrule(lr){10-13}
 & R@10 & N@10 & R@20 & N@20 & R@10 & N@10 & R@20 & N@20 & R@10 & N@10 & R@20 & N@20 \\
 \midrule
 LightGCN \cite{he2020lightgcn} & 0.1339 & 0.0803 & 0.1903 & 0.0959 & 0.0762 & 0.0424 & 0.1023 & 0.0513 & 0.0589 & 0.0393 & 0.0912 & 0.0483 \\
 SGL \cite{wu2021self} & 0.1441 & 0.0837 & 0.2043 & 0.0993 & 0.0821 & 0.0447 & 0.1195 & 0.0546 & 0.0673 & 0.0411 & 0.1071 & 0.0524 \\
 SVD-GCN \cite{peng2022svd} & 0.1773 & 0.1028 & 0.2532 & 0.1223 & 0.1082 & 0.0602 & 0.1427 & 0.0704 & 0.0816 & 0.0492 & 0.1253 & 0.0632 \\
 \midrule
 HGCF \cite{sun2021hgcf} & 0.1424 & 0.0864 & 0.2012 & 0.1043 & 0.0847 & 0.0462 & 0.1145 & 0.0568 & 0.0653 & 0.0437 & 0.1013 & 0.0521 \\
 HICF \cite{yang2022hicf} & 0.1496 & 0.0943 & 0.2087 & 0.1124 & 0.0903 & 0.0509 & 0.1217 & 0.0621 & 0.0709 & 0.0479 & 0.1089 & 0.0567 \\
 \midrule
 KGCN \cite{wang2019knowledge} & 0.1233 & 0.0721 & 0.1772 & 0.0882 & 0.0687 & 0.0362 & 0.0942 & 0.0464 & 0.0436 & 0.0251 & 0.0735 & 0.0338 \\
 KGAT \cite{wang2019kgat} & 0.1347 & 0.0815 & 0.1917 & 0.0974 & 0.0785 & 0.0431 & 0.1034 & 0.0527 & 0.0586 & 0.0401 & 0.0917 & 0.0486 \\
 KGIN \cite{wang2021learning} & 0.1437 & 0.0889 & 0.2097 & 0.1116 & 0.0895 & 0.0495 & 0.1207 & 0.0602 & 0.0676 & 0.0461 & 0.0997 & 0.0546 \\
 KGCL \cite{yang2022knowledge} & 0.1593 & 0.0957 & 0.2296 & 0.1178 & 0.0997 & 0.0543 & 0.1324 & 0.0665 & 0.0754 & 0.0536 & 0.1174 & 0.0588 \\
 \midrule
 LKGR \cite{chen2022modeling} & 0.1662 & 0.1025 & 0.2547 & 0.1242 & 0.1141 & 0.0681 & 0.1543 & 0.0791 & 0.0832 & 0.0557 & 0.1334 & 0.0679 \\
 HCMKR \cite{sun2024hyperbolic} & \underline{0.1793} & \underline{0.1073} & \underline{0.2612} & \underline{0.1263} & \underline{0.1231} & \underline{0.0689} & \underline{0.1674} & \underline{0.0804} & \underline{0.0957} & \underline{0.0571} & \underline{0.1493} & \underline{0.0726} \\
 \midrule
 \rowcolor{gray!20} SPARK(\textbf{Ours}) & \textbf{0.1987*} & \textbf{0.1154*} & \textbf{0.2791*} & \textbf{0.1347*} & \textbf{0.1293*} & \textbf{0.0753*} & \textbf{0.1823*} & \textbf{0.0864*} & \textbf{0.1016*} & \textbf{0.0624*} & \textbf{0.1575*} & \textbf{0.0779*} \\
 \midrule
 Improv. (\%) & 10.8\% & 6.9\% & 6.9\% & 6.7\% & 5.0\% & 9.3\% & 8.9\% & 7.5\% & 6.2\% & 9.3\% & 5.5\% & 7.3\% \\
 \bottomrule
\end{tabular}
\end{table*}

\subsection{Overall Performance (RQ1)}
\label{subsec:overall_performance_spark}

To answer \textbf{RQ1}, we conducted a comprehensive comparison of SPARK against a suite of baseline models on three real-world datasets. The overall performance results, presented in Table~\ref{tab:overall_results_spark}, unequivocally demonstrate SPARK's superiority. We derive the following key observations:

SPARK consistently achieves state-of-the-art (SOTA) performance across all datasets and evaluation metrics (Recall, NDCG for N=10, 20), with improvements often being statistically significant ($p < 0.05$) over the strongest baselines. This robust superiority stems from several synergistic factors inherent in SPARK's design:

\textbf{First, effective KG utilization and noise mitigation:} SPARK initiates its pipeline with a principled KG processing step. The Tucker-based scoring and representation learning (Section~\ref{subsec:kg_processing_tucker_balanced}) allows SPARK to robustly extract valuable semantic information by modeling complex multi-way interactions within KG triples. This low-rank decomposition approach inherently helps in denoising the KG and generating more compact, resilient entity embeddings, which is particularly beneficial for downstream recommendation tasks often plagued by noisy or incomplete KG data. This contrasts with simpler KG-aware models that might directly incorporate noisy KG structures.

\textbf{Second, advanced hybrid geometric modeling:} Unlike models confined to a single geometric space, SPARK's hybrid architecture (Section~\ref{subsec:hybrid_gnn_hygrec_style_full_balanced}) operates concurrently in both Euclidean and Hyperbolic (Lorentz model) spaces, providing greater representational power. The Euclidean pathway, bootstrapped by SVD-initialized features capturing global collaborative trends, effectively models homophilic relationships. Simultaneously, the Hyperbolic pathway is strategically leveraged for its aptitude in modeling hierarchical structures and power-law distributions often present in KGs or user-item interaction patterns. This dual-space approach, particularly the hyperbolic component, enhances the semantic understanding and representation quality of items, especially sparse, long-tail entities whose relational context might be better captured in a non-Euclidean geometry.

\textbf{Third, synergistic information fusion and alignment:} SPARK employs sophisticated mechanisms to integrate diverse information sources. The integration of contrastive learning (Section~\ref{subsec:contrastive_alignment_balanced}) ensures better alignment and semantic consistency between the SVD-derived collaborative view and the KG-enhanced semantic view, leading to more robust and generalizable item representations. Furthermore, the item popularity-aware adaptive fusion mechanism (Section~\ref{subsec:adaptive_fusion_balanced_concise}) dynamically tailors the final item representations. This allows SPARK to optimally leverage different information sources—collaborative signals, KG-derived semantics, and insights from different geometric spaces—for both mainstream and long-tail items, leading to more precise and personalized recommendations. This adaptability is a key differentiator from models with static fusion strategies.

When comparing SPARK to specific categories of baselines, its advantages become more apparent. Traditional KG-unaware GNNs (e.g., LightGCN, SVD-GCN) and even some basic KG-aware models (e.g., KGCN) are significantly outperformed. This is attributable to SPARK's more sophisticated integration and modeling of KG semantics, its principled noise reduction, and its exploration of non-Euclidean geometry. While hyperbolic-specific models (e.g., HGCF, LKGR, HCMKR) demonstrate the benefits of hyperbolic space, SPARK surpasses them by not solely relying on one geometric prior. Instead, its hybrid approach—combining robust SVD initialization for global collaborative signals, explicit KG modeling via Tucker decomposition for semantic richness, and adaptive fusion for tailored representations—provides a more comprehensive, flexible, and ultimately more effective framework.

The significant improvements, as quantified by the "Improv. (\%)" row in Table~\ref{tab:overall_results_spark} (e.g., up to 10.8\% in R@10 on Amazon-Book and 6.2\% on Yelp2018 over the best performing baselines), underscore the practical efficacy of SPARK's holistic design. These results strongly suggest that concurrently addressing KG quality, leveraging appropriate geometric inductive biases for diverse data characteristics, and intelligently fusing multi-source information are crucial for advancing knowledge-aware recommendation systems.

\begin{table*}[h] 
\centering
\small
\caption{Ablation study results on three datasets. The \textbf{bolded numbers} denote the most significant performance degradation when removing each component, indicating its high importance.}
\label{tab:ablation_results_spark}
\renewcommand{\arraystretch}{1.1}
\setlength{\tabcolsep}{1.8mm}
\begin{tabular}{lccccccccccccc}
\toprule
\multirow{2}{*}{Model} & \multicolumn{4}{c}{Amazon-Book} & \multicolumn{4}{c}{Alibaba-iFashion} & \multicolumn{4}{c}{Yelp2018} \\
\cmidrule(r){2-5} \cmidrule(r){6-9} \cmidrule(r){10-13}
& R@10 & N@10 & R@20 & N@20 & R@10 & N@10 & R@20 & N@20 & R@10 & N@10 & R@20 & N@20 \\
\midrule
SPARK (Full) & 0.1987 & 0.1154 & 0.2791 & 0.1347 & 0.1293 & 0.0753 & 0.1823 & 0.0864 & 0.1016 & 0.0624 & 0.1575 & 0.0779 \\
\midrule
SPARK w/o Tucker & 0.1923 & 0.1105 & 0.2724 & 0.1289 & 0.1247 & 0.0716 & 0.1765 & 0.0821 & 0.0973 & 0.0589 & 0.1502 & 0.0732 \\
SPARK w/o HypGeom & \textbf{0.1845} & \textbf{0.1067} & \textbf{0.2623} & \textbf{0.1245} & \textbf{0.1201} & \textbf{0.0689} & \textbf{0.1687} & \textbf{0.0785} & \textbf{0.0924} & \textbf{0.0551} & \textbf{0.1431} & \textbf{0.0698} \\
SPARK w/o CL & 0.1901 & 0.1089 & 0.2698 & 0.1276 & 0.1235 & 0.0704 & 0.1751 & 0.0809 & 0.0956 & 0.0571 & 0.1486 & 0.0719 \\
SPARK w/o PopFus & 0.1959 & 0.1126 & 0.2724 & 0.1321 & 0.1224 & 0.0702 & 0.1718 & 0.0803 & 0.0948 & 0.0568 & 0.1467 & 0.0721 \\
SPARK w/o SVDInit & 0.1856 & 0.1074 & 0.2641 & 0.1261 & 0.1213 & 0.0698 & 0.1704 & 0.0798 & 0.0941 & 0.0564 & 0.1459 & 0.0714 \\
\bottomrule
\end{tabular}
\end{table*}


\subsection{Ablation Study (RQ2)}
\label{subsec:ablation_study_spark}
To answer \textbf{RQ2} and rigorously evaluate the individual contribution of each key component within our SPARK framework, we conducted a comprehensive ablation study. We systematically removed core modules one by one—low-rank Tucker decomposition for KG processing (w/o Tucker), SVD-based initialization (w/o SVDInit), the hyperbolic geometry pathway (w/o HypGeom), contrastive learning for view alignment (w/o CL), and the popularity-aware adaptive fusion (w/o PopFus)—and observed the impact on performance. To create these variants, we implemented the ablations as follows: 1). \textbf{w/o Tucker} replaces the Tucker-based KG processing with a simpler translational scoring model (TransE-style). 2). \textbf{w/o SVDInit} omits the SVD-based initialization, using standard Xavier initialization for GNN features instead. 3). \textbf{w/o HypGeom} removes the entire hyperbolic GNN pathway, with predictions relying solely on the Euclidean branch. 4). \textbf{w/o CL} disables the contrastive loss term from the final objective function. 
Finally, 5). \textbf{w/o PopFus} replaces the dynamic popularity-aware gate with a static, simple averaging for signal fusion. The results, presented in Table~\ref{tab:ablation_results_spark}, lead to the following key observations about the importance of each component:


\textbf{The hyperbolic geometry pathway (HypGeom) is paramount to SPARK's performance.} As shown in Table~\ref{tab:ablation_results_spark}, removing HypGeom (variant: w/o HypGeom) consistently incurs the {most substantial performance degradation} across all datasets and metrics. For instance, on Amazon-Book, R@10 drops by approximately 7.15\% (from 0.1987 to 0.1845) and N@10 by 7.54\% (from 0.1154 to 0.1067). This starkly underscores the critical role of hyperbolic space, with its inherent ability to model hierarchical and tree-like structures with low distortion (as detailed in Section~\ref{subsec:hybrid_gnn_hygrec_style_full_balanced}), in capturing complex relationships essential for effective recommendation, especially in knowledge-rich environments. This finding strongly validates our hybrid geometric design.

\textbf{SVD-based initialization (SVDInit) provides a crucial strong inductive bias.} The variant w/o SVDInit also exhibits significant performance drops (e.g., on Amazon-Book, R@10 decreases by about 6.59\% and N@10 by 6.93\%). This highlights the importance of initializing GNN pathways with rich global collaborative priors derived from SVD. Without this strong initialization, which effectively captures overarching user-item interaction patterns (Section~\ref{subsec:hybrid_gnn_hygrec_style_full_balanced}), the model struggles to converge to optimal representations, especially in sparser datasets.

\textbf{Contrastive learning (CL) for view alignment is essential for robust and coherent representations.} Eliminating CL (w/o CL) results in a consistent and notable decline in performance (e.g., on Amazon-Book, R@10 by approx. 4.33\% and N@10 by 5.63\%). This demonstrates that explicitly aligning the SVD-collaborative view and the KG-enhanced semantic view (Section~\ref{subsec:contrastive_alignment_balanced}) is crucial for creating multi-faceted item representations that are not only semantically consistent but also more generalizable and robust to noise from individual views.

\textbf{Principled KG processing via Tucker decomposition (Tucker) significantly contributes to leveraging KG semantics effectively.} Removing the Tucker-based KG processing (w/o Tucker) leads to clear performance degradation (e.g., on Amazon-Book, R@10 by approx. 3.22\% and N@10 by 4.25\%). This confirms that effective KG denoising and the generation of robust initial entity representations via Tucker decomposition (Section~\ref{subsec:kg_processing_tucker_balanced}) are vital for successfully incorporating valuable semantic information from KGs, particularly for improving the understanding of long-tail items that rely more heavily on KG context.

\textbf{Popularity-aware adaptive fusion (PopFus) offers valuable nuanced modeling, though its relative impact varies.} While removing PopFus (w/o PopFus) also degrades performance (e.g., on Amazon-Book, R@10 by approx. 1.41\% and N@10 by 2.43\%), its impact appears less dominant than HypGeom or SVDInit in these ablation settings. Nevertheless, its contribution in dynamically tailoring the fusion of collaborative, KG, and geometric signals based on item popularity remains important for achieving fine-grained modeling and catering effectively to both mainstream and long-tail items, preventing a one-size-fits-all fusion approach.


In summary, these ablation studies confirm the significant contribution of each core component in SPARK. The results particularly emphasize the indispensable roles of the hyperbolic geometry pathway for capturing complex structures and SVD initialization for providing a strong collaborative prior. These are followed by the substantial benefits derived from contrastive learning for view coherence and Tucker-based KG processing for semantic enrichment and denoising. The popularity-aware fusion, while having a comparatively smaller direct impact in these specific ablations, complements the framework by enabling adaptive and nuanced modeling across diverse item popularities. This comprehensive analysis validates the rationality and effectiveness of SPARK's carefully designed multi-stage architecture.


\begin{figure}[t] 
    \centering
    \includegraphics[width=0.32\columnwidth]{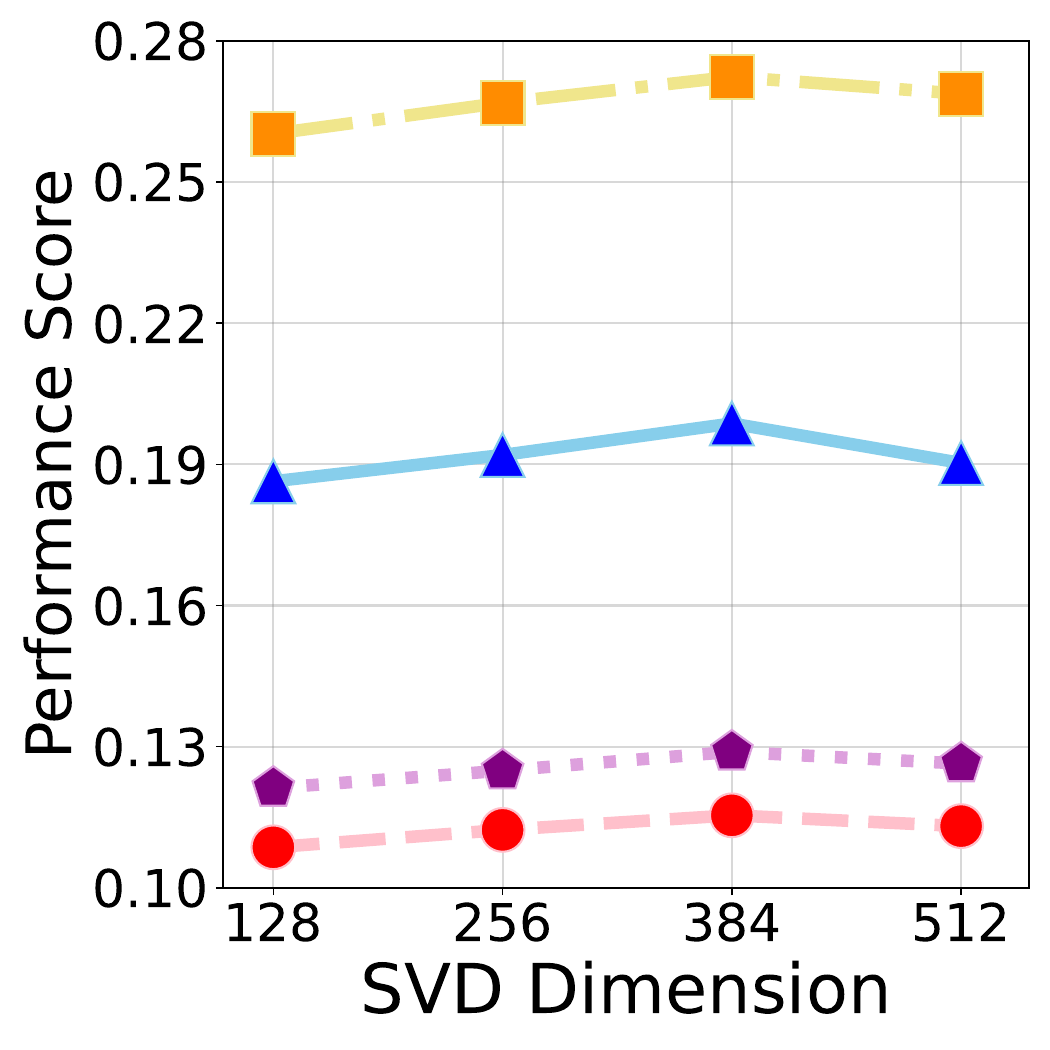}
    \hspace{-2mm} 
    \includegraphics[width=0.32\columnwidth]{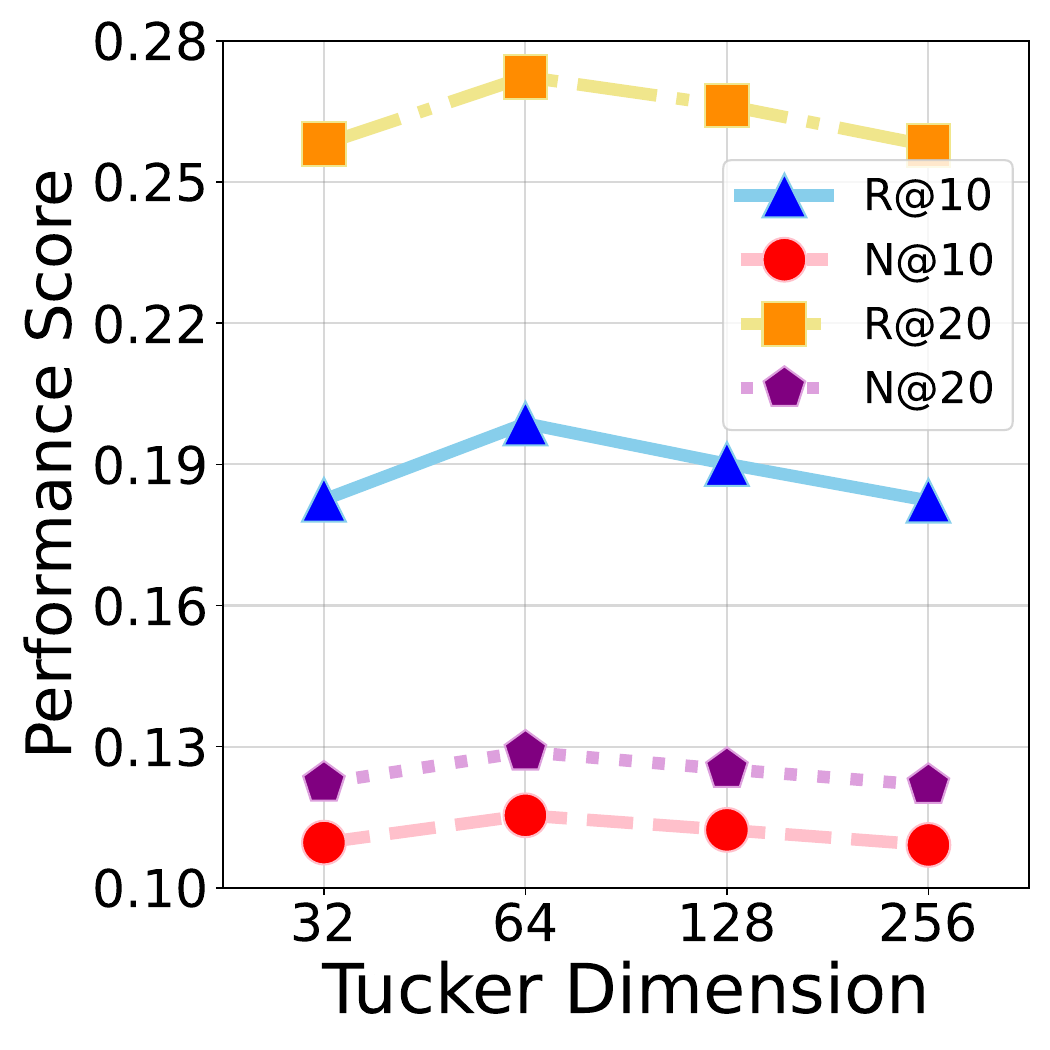}
    \hspace{-2mm}
    \includegraphics[width=0.32\columnwidth]{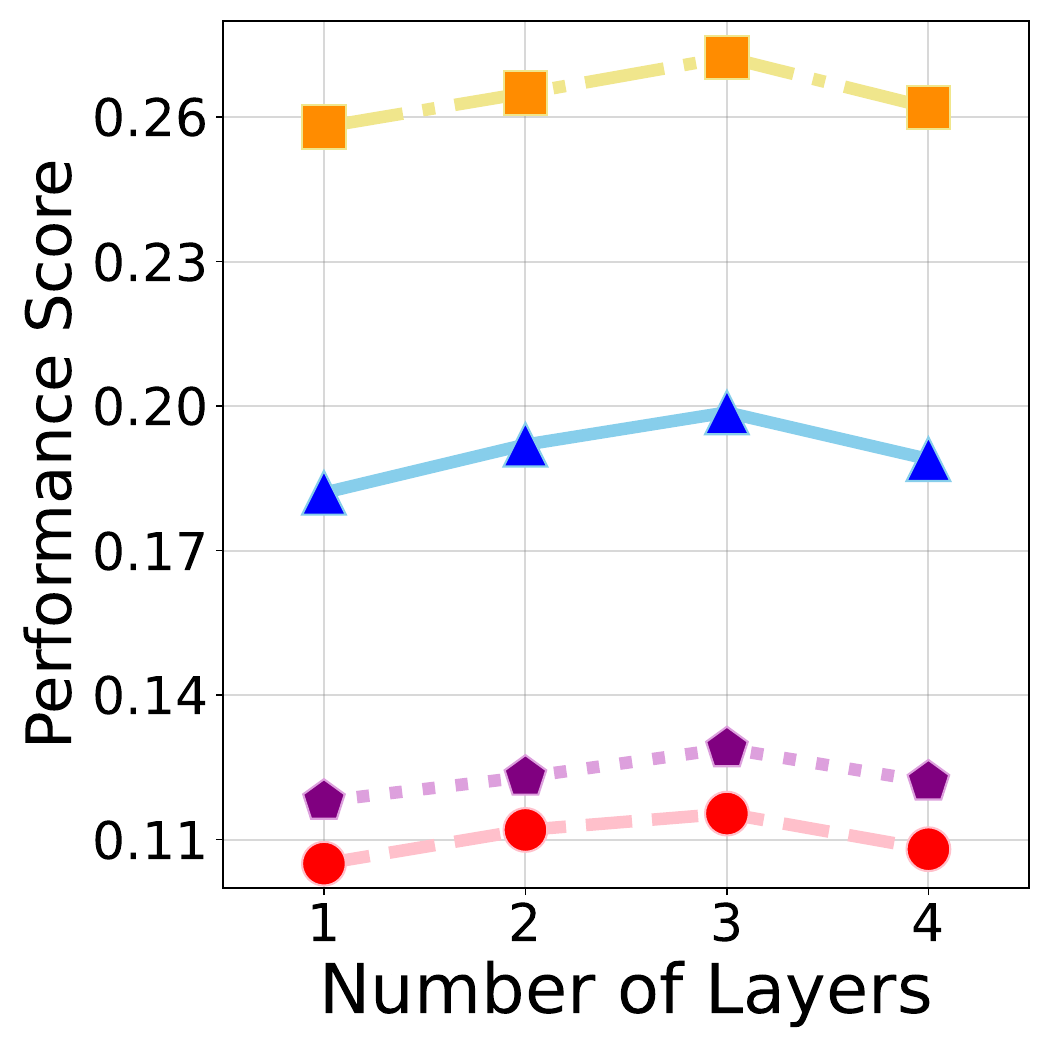}


    \vspace{-3mm} 
    \caption{
        Hyperparameter sensitivity on Amazon-Book for (left) SVD dimension $d_e$, (center) Tucker core dimension $d_c$, and (right) GNN layers $L$. R@k and N@k denote Recall@k and NDCG@k.
    }
    \label{fig:spark_sensitivity_minipage}
    \vspace{-4mm} 
\end{figure}


\subsection{Hyperparameter Sensitivity Analysis (RQ3)}
\label{subsec:hyperparam_sensitivity_spark}

To understand how different hyperparameter choices affect SPARK's performance and to provide insights for practical deployment (\textbf{RQ3}), we conducted a sensitivity analysis on several critical architectural parameters. We varied one hyperparameter at a time while keeping others fixed at their empirically determined optimal values. The results, primarily on the Amazon-Book dataset (chosen for its representative nature), are illustrated in Figure~\ref{fig:spark_sensitivity_minipage}.

\textbf{SVD Embedding Dimension ($d_e$):} As shown in Figure~\ref{fig:spark_sensitivity_minipage}(a), the SVD embedding dimension $d_e$, which determines the dimensionality of the initial global collaborative features, significantly impacts performance. We observe that optimal performance across metrics like R@10 and N@10 is achieved when $d_e=384$. Smaller dimensions (e.g., 128, 256) appear insufficient to capture the full richness of global collaborative patterns present in the user-item interaction matrix, leading to underfitting. Conversely, excessively large dimensions (e.g., 512) may introduce noise from the SVD process or make the initial projection prone to overfitting, subsequently degrading the quality of features fed into the GNN pathways. Thus, $d_e=384$ strikes an effective balance between representational capacity and model parsimony for the SVD-derived priors.

\textbf{Tucker Core Tensor Dimension ($d_c$):} The dimensionality of the Tucker core tensor, $d_c$, used in KG processing (Section~\ref{subsec:kg_processing_tucker_balanced}), is crucial for balancing semantic preservation and noise reduction. Figure~\ref{fig:spark_sensitivity_minipage}(b) reveals a distinct inverted-V pattern, with $d_c=64$ yielding peak performance. Dimensions smaller than 64 (e.g., 32) likely result in excessive compression of KG information, leading to the loss of essential semantic details. On the other hand, larger dimensions (e.g., 128, 256) may retain too much noise from the original KG or make the Tucker decomposition less effective at identifying the most salient multi-linear relationships, thereby undermining its denoising benefits. The results suggest that $d_c=64$ offers an optimal trade-off for robust KG representation learning within SPARK.

\textbf{Number of GNN Layers ($L$):} The depth of the GNN pathways, $L$, determines the extent of neighborhood aggregation. As depicted in Figure~\ref{fig:spark_sensitivity_minipage}(c), $L=3$ GNN layers generally achieve optimal performance. Using fewer layers (e.g., $L=1$ or $L=2$) may not allow the model to capture sufficiently complex higher-order relationships and propagate information effectively across the graph. Conversely, employing more layers (e.g., $L=4$) can lead to issues like over-smoothing, where node representations become indistinguishable, or overfitting, especially in sparser datasets. SPARK demonstrates relative robustness to moderate changes around $L=3$, indicating a stable balance between representational power derived from multi-hop aggregation and the risk of performance degradation due to excessive depth of networks.

These analyses collectively underscore the critical importance of judicious hyperparameter selection for realizing the full potential of SPARK. Our findings not only pinpoint optimal ranges for key architectural choices like SVD embedding dimension ($d_e=384$), Tucker core dimension ($d_c=64$), and GNN depth ($L=3$), but also illuminate the underlying trade-offs involved—such as balancing representational capacity against noise sensitivity, or aggregation depth against over-smoothing. Encouragingly, SPARK exhibits reasonable performance stability around these optima, suggesting a degree of robustness. This understanding provides valuable, practical guidance for researchers and practitioners aiming to configure SPARK efficiently and effectively for diverse datasets and recommendation scenarios, ensuring that its sophisticated mechanisms are well-calibrated to deliver superior performance.



\subsection{Case Study (RQ4)}
\label{subsec:long_tail_case_study}

\begin{figure}[!htbp]
    \centering
    \includegraphics[width=0.32\columnwidth]{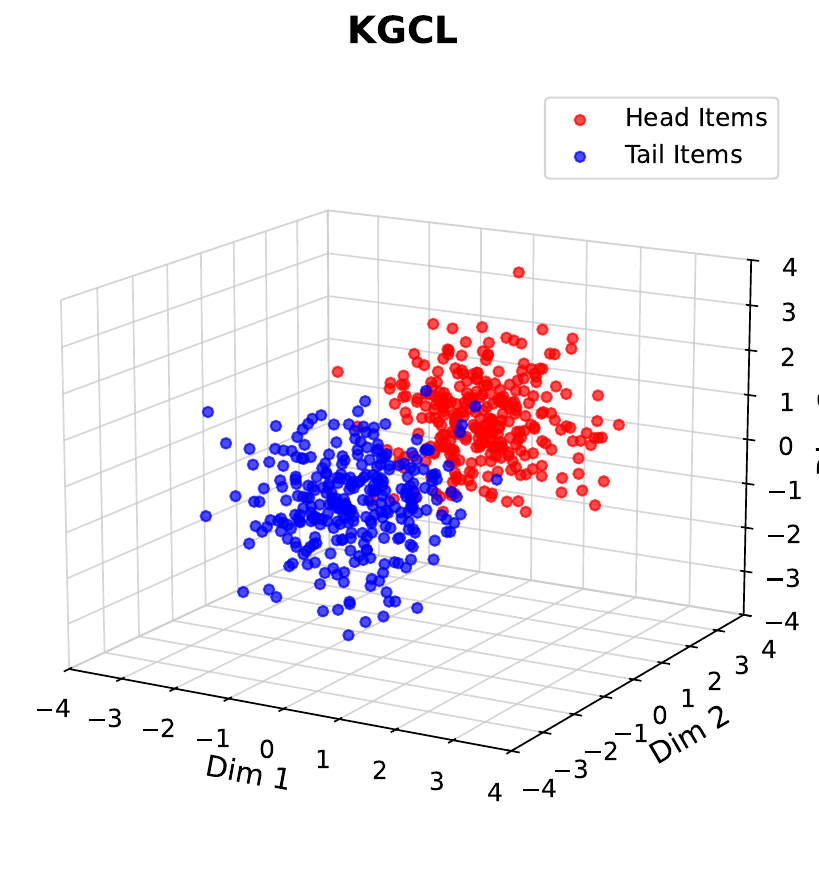}
    \hspace{-1mm} 
    \includegraphics[width=0.32\columnwidth]{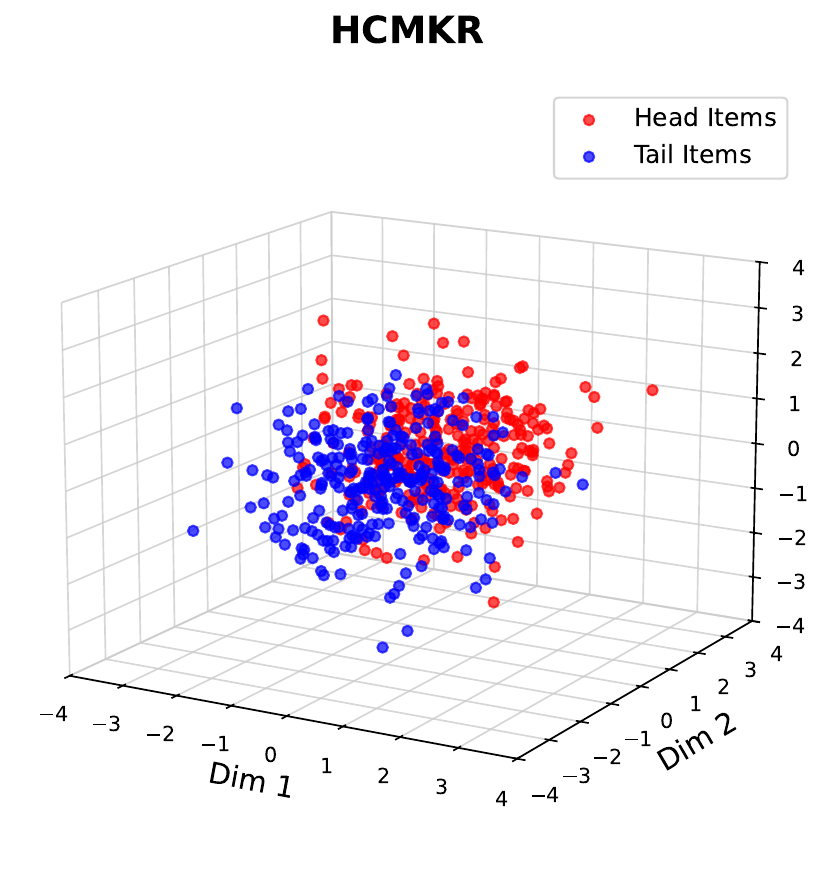}
    \hspace{-1mm}
    \includegraphics[width=0.32\columnwidth]{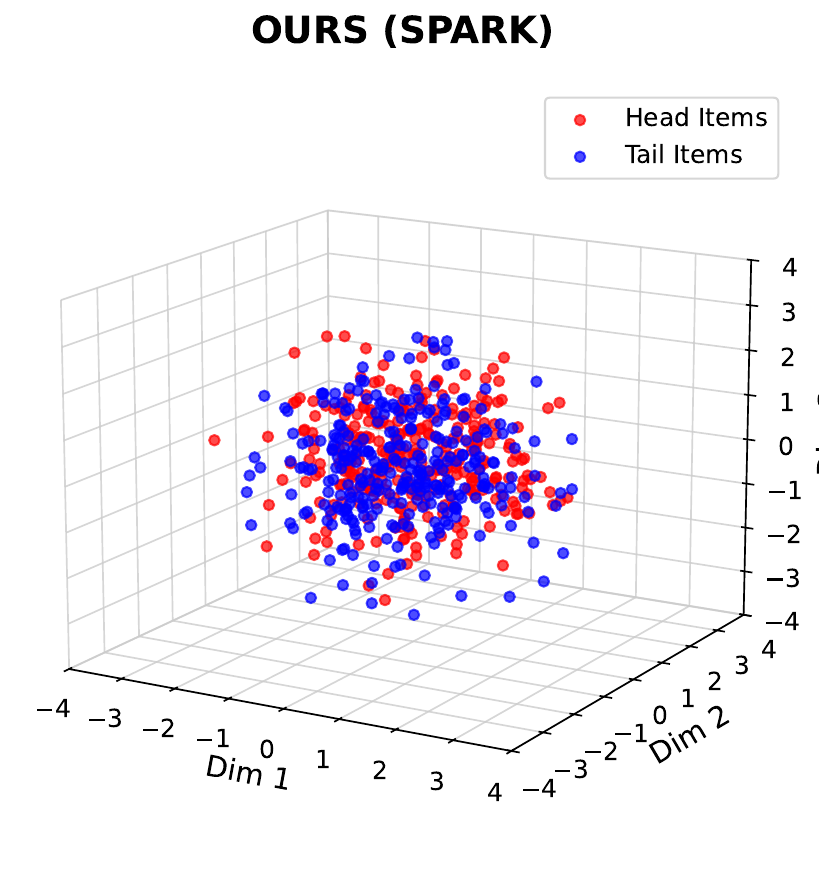}

    \makebox[0.32\columnwidth]{(a) KGCL}
    \makebox[0.32\columnwidth]{(b) HCMKR}
    \makebox[0.32\columnwidth]{(c) OURS (SPARK)}
    
    \caption{
        t-SNE visualization of learned item embeddings on Yelp2018. 
        Head items (top 10\%, blue) and tail items (bottom 10\%, red) are shown. 
        SPARK (c) learns a more balanced embedding space where tail items are better integrated with head items, compared to baselines KGCL (a) and HCMKR (b).
    }
    \label{fig:longtail_embedding_analysis_mp}
\end{figure}

A critical challenge in recommender systems is the effective representation and recommendation of long-tail items, which suffer from data sparsity. To investigate \textbf{RQ4}—how effectively SPARK addresses this problem—we conducted a primarily qualitative analysis of its learned item embeddings, focusing on the distributional disparity between popular (head) and less popular (long-tail) items as visualized in Figure~\ref{fig:longtail_embedding_analysis_mp}.

We categorized items from the Yelp2018 dataset into three popularity groups based on their historical interaction frequency: \textbf{Head} items (top 10\% most popular, depicted in blue in Figure~\ref{fig:longtail_embedding_analysis_mp}) and \textbf{Tail} items (bottom 10\% least popular, depicted in red). We then visualized the learned item embeddings (projected to 3D using t-SNE for illustration) for these head and tail items from SPARK and representative baselines (KGCL and HCMKR).

Figure~\ref{fig:longtail_embedding_analysis_mp} provides compelling visual insights into how different models represent head versus tail items. In the visualization for KGCL (Figure~\ref{fig:longtail_embedding_analysis_mp}(a)), a baseline KG-aware method, we observe a rather distinct separation between the clusters of head (blue) and tail (red) items. This segregation suggests that tail items might be learned as semantically distant from head items, potentially hindering their discovery or leading to recommendations biased towards popular items. The HCMKR model (Figure~\ref{fig:longtail_embedding_analysis_mp}(b)), which incorporates hyperbolic geometry, shows a more integrated embedding space compared to KGCL. While the separation is less pronounced, the tail items still appear to form a somewhat distinct, albeit more diffuse, grouping relative to the denser head item cluster.

In stark contrast, the visualization for OURS (SPARK) (Figure~\ref{fig:longtail_embedding_analysis_mp}(c)) reveals a significantly more balanced and discriminative embedding space. The red (tail) and blue (head) points are much more intermingled, indicating that SPARK learns representations where tail items are not isolated but are instead situated more cohesively among head items. This improved integration suggests that tail items are assigned more meaningful and comparable semantic representations relative to their popular counterparts. While distinct local clusters might still exist, the overall impression is one of a more uniform embedding space where the popularity of an item does not dictate its gross position or cause it to be pushed to a sparse, peripheral region. This visual evidence points towards SPARK's enhanced capability to model long-tail items effectively.

The improved distribution observed in SPARK's embeddings (Figure~\ref{fig:longtail_embedding_analysis_mp}(c)) can be attributed to its synergistic components. The {hybrid geometric modeling}, particularly the hyperbolic pathway, likely aids in capturing nuanced relationships that define niche tail items, preventing them from being represented as simple outliers. Furthermore, the {Tucker-based KG denoising} ensures that the semantic information used to represent these often data-sparse tail items is robust and of high quality. Crucially, the {popularity-aware adaptive fusion} mechanism allows SPARK to intelligently modulate the influence of different signals; for tail items, it can place greater emphasis on rich KG-derived semantics and potentially more expressive hyperbolic representations, rather than relying on their scarce collaborative signals. This dynamic adaptation, visually reflected in the more integrated embedding space, is key to SPARK's superior handling of long-tail items.

Collectively, the visual evidence from this case study strongly suggests that SPARK's integrated approach effectively addresses the long-tail item problem. By fostering more uniform, semantically rich, and well-integrated representations, as seen in Figure~\ref{fig:longtail_embedding_analysis_mp}(c), SPARK paves the way for more equitable and accurate recommendations, particularly enhancing the discovery of long-tail content.

\section{Related Works}
\label{sec:related-works}

\noindent\textbf{Hyperbolic Representation Learning.}
Hyperbolic geometry has gained significant traction in recommender systems for its efficacy in modeling hierarchical and power-law structures often present in user-item interactions and knowledge graphs~\cite{sun2021hgcf, yang2022hicf, chen2022modeling, sun2024hyperbolic}. Its core strength lies in embedding tree-like structures with significantly lower distortion than Euclidean space, a key feature for capturing natural item taxonomies and user preferences. 
Methods like HGCF~\cite{sun2021hgcf} and HICF~\cite{yang2022hicf} have explored hyperbolic GNNs and geometry-aware learning. Others, such as LKGR~\cite{chen2022modeling} and HCMKR~\cite{sun2024hyperbolic}, integrate hyperbolic spaces directly with KGs to capture complex entity relationships, while recent works like HDRM~\cite{yuan2025hyperbolic} even merge them with diffusion models. 
However, a primary challenge remains: these works often presuppose a clean, well-structured KG. They may not comprehensively address the issue of inherent KG noise prior to geometric modeling or offer sophisticated adaptive fusion mechanisms for signals from different geometric priors. SPARK builds upon these advancements by first employing Tucker decomposition for robust KG preprocessing—a crucial step often overlooked—and then uniquely combines geometric learning with adaptive fusion tailored to item popularity~\cite{zhao2022adaptive}.

\noindent\textbf{Knowledge-Enhanced Recommendation.}
Leveraging KGs to alleviate data sparsity and cold-start issues is a pivotal direction in recommendation~\cite{cao2019unifying, wang2019kgat, lin2023autodenoise}. Approaches broadly fall into embedding-based (e.g., CKE~\cite{zhang2016collaborative}), path-based (e.g., RippleNet~\cite{wang2018ripplenet}), and GNN-based methods (e.g., KGAT~\cite{wang2019kgat}). While GNNs excel at capturing structural semantics, most operate in Euclidean space, limiting their capacity for complex hierarchies and making them susceptible to KG noise. SPARK distinguishes itself with a holistic, multi-stage pipeline. Its Tucker-based refinement explicitly tackles noise from the outset~\cite{xu2024multi}. Subsequently, its SVD-initialized hybrid geometric GNN and popularity-aware fusion mechanism~\cite{li2023hamur, fu2025unified} provide a more nuanced and robust solution compared to prior works that typically focus on a subset of these challenges.

\noindent\textbf{Advanced Modeling Paradigms.}
Our work is also informed by broader trends in sequential and multi-modal systems. 
\textbf{Data augmentation} and denoising techniques aim to enrich sparse user data. 
Examples include automated instance filtering~\cite{lin2023autodenoise} and the development of efficient sequential architectures like sparse Transformers~\cite{li2023strec} and pure MLP models~\cite{gao2024smlp4rec}.
\textbf{Reinforcement Learning (RL)} frameworks optimize for long-term engagement, tackling challenges like vast action spaces~\cite{liu2023exploration} with support from user simulators~\cite{zhao2023kuaisim, zhao2018deep, zhao2018recommendations, zhao2020whole}. 
Furthermore, \textbf{Large Language Models (LLMs)} are revolutionizing the field by interpreting heterogeneous information. 
LLMs are now applied to tasks from multi-domain recommendation~\cite{zhang2024m3oe, wang2025behavior, fu2025unified} and medication recommendation~\cite{liu2024large} to KG completion and QA~\cite{xu2024multi, xu2025harnessing, chen2024large}. 
These approaches highlight a shared goal: building intelligent and robust recommender systems, to which SPARK contributes a novel, geometry-centric perspective.

\section{Conclusion}
\label{sec:conclusion}

This paper introduced SPARK, a novel multi-stage framework that effectively tackles key challenges in knowledge-aware recommendation, including knowledge graph (KG) noise, geometric limitations, and the nuanced modeling of mainstream versus long-tail items. 
SPARK's efficacy stems from its synergistic pipeline: low-rank Tucker decomposition refines KG inputs, a hybrid geometric GNN learns diverse representations, and a popularity-aware adaptive fusion mechanism, further enhanced by contrastive alignment, optimally combines these signals. Extensive experiments on three real-world datasets demonstrate SPARK's significant performance improvements over state-of-the-art baselines, particularly in enhancing long-tail recommendations. This research paves the way for more robust, adaptable, and geometrically-informed knowledge-aware recommender systems. Future work could explore extending the SPARK framework to other challenging scenarios, such as cold-start recommendation, where the denoised KG representations could provide crucial semantic priors. Furthermore, adapting the fusion mechanism to incorporate temporal dynamics presents another promising direction. The modular and principled design of SPARK provides a solid foundation for these future explorations.


\begin{acks}
This research was partially supported by National Natural Science Foundation of China (No.62502404), Hong Kong Research Grants Council's Research Impact Fund (No.R1015-23), Collaborative Research Fund (No.C1043-24GF), General Research Fund (No.11218325), Institute of Digital Medicine of City University of Hong Kong (No.9229503), Huawei (Huawei Innovation Research Program), Tencent (CCF-Tencent Open Fund, Tencent Rhino-Bird Focused Research Program), Alibaba (CCF-Alimama Tech Kangaroo Fund No. 2024002), Ant Group (CCF-Ant Research Fund), Didi (CCF-Didi Gaia Scholars Research Fund), Kuaishou, and Bytedance.
\end{acks}

\newpage

\section*{GenAI Usage Disclosure}
We provide full disclosure of our use of GenAI tools throughout this research. We utilized Claude (Anthropic) for language enhancement, including grammar checking, sentence structure improvement, and academic expression refinement. As non-native English speakers, the AI tool was employed solely to improve the linguistic quality of our original research without generating or altering the core research content, methodology, or findings. All conceptual frameworks, experimental designs, and research conclusions are entirely our own intellectual contributions.

\balance
\bibliographystyle{ACM-Reference-Format}
\balance
\bibliography{sample-base}



\end{document}